\documentclass[aps,prev,twocolumn,preprintnumbers,floatfix,nofootinbib]{revtex4-1}
\usepackage[latin1]{inputenc}
\usepackage{amssymb}
\usepackage{color}
\usepackage{graphicx}
\usepackage{dcolumn}
\usepackage{bm}
\usepackage[header,title,page,titletoc]{appendix}
\usepackage{amsmath}
\usepackage{amsfonts}%
\providecommand{\U}[1]{\protect \rule{.1in}{.1in}}
\begin{document}
%\title{Nonuniversality of the neutral interactions in the $331_{RHN}$}
%\title{CP violation in the neutral mesons systems in the 331}
%\title{Novel CP violations sources in the 331}
\title{Novel sources of Flavor Changed Neutral Currents in the $331_{RHN}$ model}
\author{D. Cogollo}
\email{diegocogollo@df.ufcg.edu.br}
\affiliation{Departamento de
Fisica, Universidade Federal de Campina Grande, Caixa Postal 10071, 58109-970 Campina Grande, Paraiba, Brazil}
\author{F.S. Queiroz}
\email{fqueiroz@fnal.gov}
\affiliation{Center for Particle Astrophysics, Fermi National Accelerator Laboratory, Batavia, IL 60510, USA}
\affiliation{Departamento de Fisica, Universidade Federal da Paraiba, Caixa Postal 5008, 58051-970, Joao Pessoa, PB, Brazil}
\author{P.R. Teles}
\email{patricia.teles@ufabc.edu.br}
\affiliation{Centro de Ci\^encias Naturais e Humanas, Universidade Federal do ABC, R. Santa Adlia 166, 09210-170, Santo Andr\'e - SP, Brazil.}
\author{A. Vital de Andrade}
\email{aubery.vital@df.ufcg.edu.br}
\affiliation{Departamento de
Fisica, Universidade Federal de Campina Grande, Caixa Postal 10071, 58109-970 Campina Grande, Paraiba, Brazil}

\begin{abstract}
Sources of Flavor Changed Neutral Currents (FCNC) emerge naturally from a well motivated framework called 3-3-1 with right-handed neutrinos model, $331_{RHN}$ for short, mediated by an extra neutral gauge boson $Z^{\prime}$. Following previous works we calculate these sources and in addition we derive new ones coming from CP-even and -odd neutral scalars which appear due to their non-diagonal interactions with the physical standard quarks. Furthermore, by using 4 texture zeros for the quark mass matrices, we derive the mass difference terms for the neutral mesons systems $K^{0}-\bar{K}^{0}$, $D^{0}-\bar{D}^{0}$ and $B^0-\bar{B^0}$ and show that, though one can discern that the $Z^{\prime}$ contribution is the most relevant one for mesons oscillations purposes, scalars contributions play a role also in this processes and hence it is worthwhile to investigate them and derive new bounds on space of parameters. In particular, studying the $B^0-\bar{B^0}$ system we set the bounds $M_{{Z^\prime}} \gtrsim 4.2$~TeV and  $M_{S_2},M_{I_3} \gtrsim 7.5$ TeV  in order to be consistent with the current measurements.
\end{abstract}



\maketitle

\section{Introduction}

The accurate measurements of processes involving neutral mesons oscillations like $K^{0}-\bar{K}^{0}$,$D^{0}-\bar{D}^{0}$ and $B^{0}_d-\bar{B^0_d}$ have become a great laboratory to check Standard Model (SM) consistency as well as 
new physics models feasibilities. It is well known that mesons oscillations  are forbidden
at tree level in the SM, but may be generated by taking into account loop corrections or introducing
dimension-6 operators~\cite{dimensio6operators}. Experimental precision data up to now are completely consistent with SM predictions regarding mesons oscillations, thus several analysis have been performed in order to test and constrain new models. In the case of 331 models ~\cite{331-1,331-2}, FCNC arise naturally at tree level in the quark sector because the new neutral gauge boson $Z^{\prime}$ predicted by the model couples differently with the third family, leading to non-universal interactions. Up to now it has been thought that this gauge boson was the unique source of FCNC at tree level in the 331 model. Here, we are going to show that there are two other sources induced by CP-even and -odd scalars instead.  In this work will be adopting a 4 texture zeros for the quark mass matrices in the calculation of mass difference terms for the neutral mesons systems $K^{0}-\bar{K}^{0}$, $D^{0}-\bar{D}^{0}$ and $B^0-\bar{B^0}$, hence different conclusions may arise in more general approaches. In order to explain these novel contributions we are going to put them in perspective, firstly summarizing the key features of the model, and then describing how the scalars interact with the SM quarks.

\section{THE MODEL}

Our framework is the  $331_{RHN}$ model ~\cite{331rhn,331rhn2}, based in the gauge group $SU(3)_{c}\bigotimes SU(3)_{L}\bigotimes U(1)_{N}$, a direct extension of the electroweak sector of the SM. This model features many of the SM virtues while elegantly providing interesting explanations for dark matter signals~\cite{331DM} and for many theoretical questions, such as number of families~\cite{meupaper1} among others \cite{others}. Since our goal is to unveil the sources of FCNC we will leave out detailed discussions about the model and give a prompt and sufficient description of it in order to allow the reader to follow our reasoning. For those who are seeking a complete description of the model we recommend the reviews ~\cite{331-1, 331-2, 331rhn, 331rhn2}.

\subsection{Fermionic content}

Likewise the SM, the leptonic sector is arranged with left-handed fields appearing in triplets, $f_{aL} = (\nu^a_L,l^a_L,(\nu^a_R)^c)^T$, and right-handed ones in singlets, $e_{a R}$, where $a=1,2,3$ represents the three generations. In the hadronic sector, the first two families are placed as anti-triplets $Q_{i L} = (d_{i L}, -u_{i L}, D^{\prime}_{i L} )^T$, with $i=1,2$,  while the third one is placed as triplet, $Q_{3 L} = (u_{3 L}, d_{3 L}, U^{\prime}_{3 L} )^T$. The first two and the third family of quarks are in different representations due to an anomaly cancellation requirement adequately described in previous works \cite{anomaly,meupaper1}. As a consequence of this non-universality in the quark sector, sources of FCNC will arise at tree level in the $331_{RHN}$ model as we will show later. Similarly to the SM, these fermions acquire mass through spontaneous symmetry breaking mechanism in the scalar sector presented hereafter.

\subsection{Scalar content}

The scalar sector is composed of three scalar triplets namely,

\begin{eqnarray}
\label{conteudoescalar}
\chi & = & (\chi^0, \chi^-, \chi^{\prime 0})^T,\nonumber \\
\rho & = & (\rho^+, \rho^0, \rho^{\prime +})^T,\nonumber \\
\eta & = & (\eta^0, \eta^-, \eta^{\prime 0})^T.
\end{eqnarray}

These scalars allow us to build the scalar potential,

\begin{eqnarray} V(\eta,\rho,\chi)&=&\mu_\chi^2 \chi^2 +\mu_\eta^2\eta^2
+\mu_\rho^2\rho^2+\lambda_1\chi^4 +\lambda_2\eta^4
\nonumber \\
&& +\lambda_3\rho^4+ \lambda_4(\chi^{\dagger}\chi)(\eta^{\dagger}\eta)
+\lambda_5(\chi^{\dagger}\chi)(\rho^{\dagger}\rho) \nonumber \\
&& +\lambda_6
(\eta^{\dagger}\eta)(\rho^{\dagger}\rho) + \lambda_7(\chi^{\dagger}\eta)(\eta^{\dagger}\chi)
\nonumber \\
&& +\lambda_8(\chi^{\dagger}\rho)(\rho^{\dagger}\chi)+\lambda_9
(\eta^{\dagger}\rho)(\rho^{\dagger}\eta)
\nonumber \\
&&-\frac{f}{\sqrt{2}}\epsilon^{ijk}\eta_i \rho_j \chi_k +\mbox{H.c}.,
\label{potential}
\end{eqnarray}

which is responsible for the spontaneous symmetry breaking mechanism $SU(3)_{C} \otimes SU(3)_{L} \otimes U(1)_{N}\rightarrow SU(3)_{C} \otimes SU(2)_{L} \otimes U(1)_{Y}$ and $SU(3)_{C} \otimes SU(2)_{L} \otimes U(1)_{Y} \rightarrow U(1)_{EM}$ as described in~\cite{331DM,darkmatter}. After the diagonalization procedure we find a CP-even $(S_{1},S_{2},H)$ and a CP-odd $(I^0_1,I^0_2,I^0_3)$ basis as follows,

\begin{equation}
\label{escalaresfisicos}
S_{1} = R_{\chi^{\prime}},\ \ S_{2} = \dfrac{1}{\sqrt{2}}(R_{\eta}-R_{\rho}),\ \ H = \dfrac{1}{\sqrt{2}}(R_{\eta}+R_{\rho}).
\end{equation}

\begin{eqnarray}
M^{2}_{S_{1}} & = & \frac{v^{2}}{4}+2v_{\chi^\prime}^{2}\lambda_{1},\nonumber \\
M^{2}_{S_{2}} & = & \frac{1}{2}(v_{\chi^\prime}^{2}+2v^{2}(2\lambda_{2}-\lambda_{6})),\nonumber \\
M^{2}_{H} & = & v^{2}(2\lambda_{2}+\lambda_{6})
\label{massashiggs}
\end{eqnarray}

\begin{equation}
\label{psescalaresfisicos}
I_{1}^{0} \sim -I_{\chi^{\prime}},\ \ I_{2}^{0} \sim \frac{1}{\sqrt{2}}(I_{\rho}-I_{\eta}),\ \ I_{3}^{0} \sim \frac{1}{\sqrt{2}}(I_{\rho}+I_{\eta}).
\end{equation}

\begin{eqnarray}
\label{escalaresfisicos2}
M^{2}_{I_{1}^{0}} = 0,\
M^{2}_{I_{2}^{0}} = 0,\
M^{2}_{I_{3}^{0}} = \frac{1}{2}(v_{\chi^\prime}^{2}+\frac{v^{2}}{2}),
\end{eqnarray} where $v$ is the vev of the neutral scalars $\rho^0$ and $\eta^0$ while $v_{\chi^\prime}$ is the vev of the neutral field  $\chi^{\prime 0}$. Here $v = v_{SM}/\sqrt{2}$.

In Eq.\eqref{escalaresfisicos} $H$ stands for the Standard Higgs boson, $S_1$ and $S_2$ are two heavy CP-even scalars. In Eq.\eqref{psescalaresfisicos} $I_{1}^{0}$ and $I_{2}^{0}$ fields are Goldstone bosons while $I_{3}^{0}$ is a heavy massive pseudoscalar. In particular the scalars $S_2$ and $I_{3}^{0}$ are responsible for the FCNC in the scalar sector of the 331 model as we will demonstrate in the next sections. The other scalars of the model such as the charged ones, are not important in our analyses.

The triplet of scalars given in the Eq.(\ref{conteudoescalar}), will be responsible for generating all fermions masses, except for neutrinos, through the Yukawa lagrangian:
\begin{eqnarray}
{-\cal{L}}^{Yuk} & = & \lambda_{2ij}\bar{Q}_{iL}\chi^{*}D^{\prime}_{jR} + \lambda_{1}\bar{Q}_{3L}\chi U^{\prime}_{3R} + \lambda_{4ia}\bar{Q}_{iL}\eta^{*}d_{aR} \nonumber\\
                & + & \lambda_{3a}\bar{Q}_{3L}\eta u_{aR}  + \lambda_{1a}\bar{Q}_{3L}\rho d_{aR} + \lambda_{2ia}\bar{Q}_{iL}\rho^{*}u_{aR}\nonumber\\
                & + & G_{aa}\bar{f}_{aL}\rho e_{aR} + H.C.
\label{yukawafermions}
\end{eqnarray}

Mass terms for neutrinos are obtained either by dimension five effective operators \cite{paulo} or by adding a scalar sextet ~\cite{scalarsextet,meu1,meu3} or a scalar anti-triplet ~\cite{331rhn}. Since the neutrinos masses are completely irrelevant to our discussions we are going to skip to the gauge sector.

\subsection{Gauge sector}

In the gauge sector the model recovers the standard gauge bosons and adds five more, known as $V^{+},V^{-},U^{0},U^{0\dagger},Z^{\prime}$; the first four carry two units of lepton number and thus are called bileptons. As we can see in Eq.(\ref{massvec}) their masses are roughly determined by the scale of symmetry breaking of the model, the value of $v_{\chi^\prime}$. In particular the new neutral gauge bosons $Z^{\prime}$ is under novel LHC experiments results regarding dilepton ressonance searches at $\sqrt{s}= 7$ TeV~\cite{limitZlinhaLHC}. Assuming that $Z^{\prime}$ and Z share the same couplings to fermions, these experiments have imposed a strong constraint on $Z^{\prime}$ mass, $M_{Z^{\prime}}\gtrsim 1.6$ TeV with 95\% C.L. CMS searches for a heavy gauge boson $W^{\prime}$ \cite{Wprime}, have put competitive bounds on the spectrum of the model. The latter provides equivalent constraints on parameter space of the model. We will take $Z^{\prime}$ for simplicity, with no impact on our conclusions. However we highlight that in the $331_{RHN}$ the couplings to fermions involving $Z^{\prime}$ are lower than the ones involving the Z, as can be checked in  Eq.\eqref{zprimau}. How precisely this experimental constraint affects our model is completely out of the scope of this work.  Notwithstanding, we are going to be conservative and adopt this lower mass limit throughout our analysis. 

\begin{eqnarray}
m_{W^\pm}^2    & = & \frac{1}{4}g^2v_{SM}^2\,,\,m^{2}_{Z} = m_{W^\pm}^2/C^{2}_{W}, \nonumber \\
m^2_{Z^\prime} & = & \frac{g^{2}}{4(3-4S_W^2)}\left[4C^{2}_{W}v_{\chi^\prime}^2 +\frac{v^{2}}{ C^{2}_{W}}+\frac{v^{2}(1-2S^{2}_{W})^2}{C^{2}_{W}}\right ],\nonumber \\
m^2_{V^\pm}    & = & \frac{1}{4}g^2(v_{\chi^\prime}^2+v^2)\,,\,m^2_{U^0}= \frac{1}{4}g^2(v_{\chi^\prime}^2+v^2).
\label{massvec}
\end{eqnarray}

So far we have described the main features of the model. The next sections will be devoted to explaining how the FCNC emerge in the $331_{RHN}$ model, and to quantify them according to recent data.

\section{Neutral currents via a gauge boson exchange}

As aforementioned FCNC is suppressed in the SM at tree level but reveals itself naturally in the $331_{RHN}$ model. In the most general case, the neutral gauge bosons of the model namely, $Z_{1}$ and $Z_{2}$ mix and provide the neutral currents derived in appendix \ref{appendix1}. However it has been shown previously that $\Phi$, the mixing angle between the physical bosons $Z_{1}$ and $Z_{2}$ is of the order $-3.979 \times 10^{-3} < \Phi < 1.309 \times 10^{-4}$ ~\cite{meupaper1}. This lets us explore the limit case $\Phi=0$, which makes $Z^1 \equiv Z$ and $Z^2 \equiv Z^{\prime}$, whose masses are given in Eq.(\ref{massvec}). Assuming this limit from now on, we may write their neutral currents with the standard quarks in the simple form,

\begin{eqnarray}
\ensuremath{\mathcal{L}}_{u+d}^{Z} & = & \frac{g}{2C_{W}}\bar{u}_{L}^{a}\gamma^{\mu}\left( \frac{3-4S_{W}^{2}}{3}\right) u_{L}^{a}Z_{\mu} \nonumber \\
                                   &   & +\frac{g}{2C_{W}}\bar{d}_{L}^{a}\gamma^{\mu}\left( \frac{2S_{W}^{2}-3}{3}\right) d_{L}^{a}Z_{\mu},
\label{cnLRquarks}
\end{eqnarray}

with $a=1,2,3$ and,

\begin{eqnarray}
\ensuremath{\mathcal{L}}^{Z^{\prime}}_{u} & = & -\frac{g}{2C_{W}}\{\bar u_{3L}\gamma^{\mu}[\frac{(3-2S_{W}^{2})}{3\sqrt{3-4S_{W}^{2}}}]u_{3L}\}Z_{\mu}^{\prime} \nonumber \\
                                          &   & + \frac{g}{2C_{\omega}}\{\bar u_{iL}\gamma^{\mu}[\frac{(3-4S_{W}^{2})}{3\sqrt{3-4S_{W}^{2}}}]u_{iL}\}Z_{\mu}^{\prime},
\label{zprimau}
\end{eqnarray}

\begin{eqnarray}
\ensuremath{\mathcal{L}}^{Z^{\prime}}_{d} & = & -\frac{g}{2C_{W}}\{\bar d_{3L}\gamma^{\mu}[\frac{(3-2S_{W}^{2})}{3\sqrt{3-4S_{W}^{2}}}]d_{3L}\}Z_{\mu}^{\prime} \nonumber \\
                                          &   & + \frac{g}{2C_{W}}\{\bar d_{iL}\gamma^{\mu}[\frac{(3-4S_{W}^{2})}{3\sqrt{3-4S_{W}^{2}}}]d_{iL}\}Z_{\mu}^{\prime},
\label{zprimad}
\end{eqnarray}

with $i=1,2$.

We may easily recognize Eq.\eqref{cnLRquarks} as the universal interaction among standard quarks and the $Z$ boson. On the other hand, it is evident from Eq.\eqref{zprimau} and Eq.\eqref{zprimad} that this is not the case for the interactions mediated by the $Z^{\prime}$ boson, because the quarks $u_1\equiv u$ and $u_2 \equiv c$ couple differently from $u_3\equiv t$ with $Z^{\prime}$. Hence we have shown that the $Z^{\prime}$ has non-universal interactions with standard quarks. Summing up the family index $\textit{a}$, Eq.\eqref{cnLRquarks} becomes,

\begin{eqnarray}
\ensuremath{\mathcal{L}}_{u+d}^{Z} & = & \frac{g}{2C_{W}}(\frac{3-4S_{W}^{2}}{3})
\begin{pmatrix}
\bar{u} & \bar{c} & \bar{t}
\end{pmatrix}_{L}
\gamma^{\mu}Z_{\mu}
\begin{pmatrix}
u\\
c\\
t
\end{pmatrix}_{L}+  \nonumber \\
                                   &    &\frac{g}{2C_{W}}(\frac{2S_{W}^{2}-3}{3})
\begin{pmatrix}
\bar{d} & \bar{s} & \bar{b}
\end{pmatrix}_{L}
\gamma^{\mu}Z_{\mu}
\begin{pmatrix}
d\\
s\\
b
\end{pmatrix}_{L}.
\label{trocadesaborneutraMP}
\end{eqnarray}

It is important to emphasize that Eq.\eqref{trocadesaborneutraMP} is written in the flavor basis. Mass eigenstates are a superposition of these flavor eigenstates and both are related by the well known transformations,
\begin{equation}
\begin{pmatrix}
u\\
c\\
t
\end{pmatrix}_{L,R}
=
V_{L,R}^{u}
\begin{pmatrix}
u^{\prime}\\
c^{\prime}\\
t^{\prime}
\end{pmatrix}_{L,R},
\begin{pmatrix}
d\\
s\\
b
\end{pmatrix}_{L,R}
=
V_{L,R}^{d}
\begin{pmatrix}
d^{\prime}\\
s^{\prime}\\
b^{\prime}
\end{pmatrix},
\label{misturaquarksMP}
\end{equation}

where $V_{L,R}^{u}$ and $V_{L,R}^{d}$ are $3\times 3$ unitary matrices which diagonalize the mass matrices for up and down standard quarks. The usual Cabibbo-Kobayashi-Maskawa (CKM) matrix is defined as $V_{CKM}=(V_{L}^{u})^{\dagger}(V_{L}^{d})$ ~\cite{cabibbo,Makoto}. Applying the transformations given in Eq.(\ref{misturaquarksMP}) on Eq.(\ref{trocadesaborneutraMP}) we obtain something proportional to,

\begin{eqnarray}
\ensuremath{\mathcal{L}}_{u+d}^{Z} & \backsim &
\begin{pmatrix}
\bar{u}^{\prime} & \bar{c}^{\prime} & \bar{t}^{\prime}
\end{pmatrix}_{L}
(V_{L}^{u})^{\dagger}(V_{L}^{u})\gamma^{\mu}Z_{\mu}
\begin{pmatrix}
u^{\prime}\\
c^{\prime}\\
t^{\prime}
\end{pmatrix}_{L} + \nonumber \\
                                  &          &\begin{pmatrix}
\bar{d}^{\prime} & \bar{s}^{\prime} & \bar{b}^{\prime}
\end{pmatrix}_{L}
(V_{L}^{d})^{\dagger}(V_{L}^{d})\gamma^{\mu}Z_{\mu}
\begin{pmatrix}
d^{\prime}\\
s^{\prime}\\
b^{\prime}
\end{pmatrix}_{L}.
\label{cu}
\end{eqnarray}

From Eq.\eqref{cu} we can verify that due to the unitarity property of these matrices, ($(V_{L}^{u})^{\dagger}(V_{L}^{u})=(V_{L}^{d})^{\dagger}(V_{L}^{d})$=1), FCNC processes are not present in the interactions mediated by the $Z$ boson. Be that as it may, the $Z^{\prime}$ boson does mediate FCNC processes at tree level, since it is not possible to write these interactions in a condensed form like Eq.\eqref{cu}, as we will clearly show below. First we may notice that Eq.\eqref{zprimau} and Eq.\eqref{zprimad} can be written as,

\begin{eqnarray}
\ensuremath{\mathcal{L}}^{Z^{\prime}}_{u}  & = & \frac{g}{2C_{W}}\left(  \frac{3-4 S_{W}^{2} }{ 3 \sqrt{ 3-4 S_{W}^{2} } }\right) \bar{u}_{aL}\gamma_\mu u_{aL} Z_{\mu}^{\prime}\nonumber\\
   &   & -\frac{g}{2C_{W}}\left(  \frac{6(1- S_{W}^{2}) }{ 3 \sqrt{ 3-4 S_{W}^{2} } }\right) \bar{u}_{3L}\gamma_\mu u_{3L} Z_{\mu}^{\prime},
   \label{zprimaufisico}
\end{eqnarray}

\begin{eqnarray}
\ensuremath{\mathcal{L}}^{Z^{\prime}}_{d}  & = & \frac{g}{2C_{W}}\left(  \frac{3-4 S_{W}^{2} }{ 3 \sqrt{ 3-4 S_{W}^{2} } }\right) \bar{d}_{aL}\gamma_\mu d_{aL} Z_{\mu}^{\prime}\nonumber\\
   &   & -\frac{g}{2C_{W}}\left(  \frac{6(1- S_{W}^{2}) }{ 3 \sqrt{ 3-4 S_{W}^{2} } }\right) \bar{d}_{3L}\gamma_\mu d_{3L} Z_{\mu}^{\prime}.
\label{zprimadfisico}
\end{eqnarray}

We can already distinctly observe that the second terms of Eqs.(\ref{zprimaufisico})-(\ref{zprimadfisico}) contribute to FCNC phenomenon at tree level. Writting Eq. \eqref{misturaquarksMP} explicitly we find:

\begin{eqnarray}
u_{aL} & = & (V_{L}^{u})_{ab}  u^{\prime}_{bL},\nonumber\\
\bar{u}_{aL} & = & \bar{u^{\prime}}_{bL} (V_{L}^{u})_{ab}^{\ast},\nonumber\\
d_{aL} & = & (V_{L}^{d})_{ab}  d^{\prime}_{bL},\nonumber\\
\bar{d}_{aL} & = & \bar{d^{\prime}}_{bL} (V_{L}^{d})_{ab}^{\ast},
\label{transformationL}
\end{eqnarray}
with $a,b=1,2,3$.

Applying Eq.\eqref{transformationL} on Eqs.\eqref{zprimaufisico}-\eqref{zprimadfisico} we obtain the Lagrangian among the physical up and down standard quarks with the $Z^{\prime}$ boson,   

\begin{equation}
\mathcal{L}^{K^{0}-\bar{K^{0}}}_{Z'}= \left(  \frac{ -g\ C_W }{  \sqrt{ 3-4 S_{W}^{2} } }\right)\{ (V^d_L)^{\ast}_{31}(V^d_L)_{32}\}[\bar{d^{\prime}_{1L} }\gamma_\mu d^{\prime}_{2L}]Z^{\prime}
\label{FCNC1}
\end{equation}

\begin{equation}
\mathcal{L}^{D^{0}-\bar{D^{0}}}_{Z'}= \left(  \frac{ -g\ C_W }{  \sqrt{ 3-4 S_{W}^{2} } }\right)\{(V^u_L)^{\ast}_{31}(V^u_L)_{32}\}[\bar{u^{\prime}_{1L} }\gamma_\mu u^{\prime}_{2L}]Z^{\prime}
\label{FCNC2}
\end{equation}

\begin{equation}
\mathcal{L}^{B^{0}_d-\bar{B^{0}_d}}_{Z'}= \left(  \frac{ -g\ C_W }{  \sqrt{ 3-4 S_{W}^{2} } }\right)\{(V^d_L)^{\ast}_{31}(V^d_L)_{33}\}[\bar{d^{\prime}_{1L} }\gamma_\mu d^{\prime}_{3L}]Z^{\prime},
\label{FCNC3}
\end{equation}

in agreement with \cite{vovan,PRDColombia}. These terms lead to the mass difference terms of the mesons system $K^{0}-\bar{K^{0}},D^{0}-\bar{D^{0}}$ and $B^{0}_d-\bar{B^{0}_d}$ respectively, as we will show further. (We are neglecting the $B^{0}_s-\bar{B^{0}_s}$ system for proving weaker constraints).

 So far we have found the known sources of FCNC which come from the $Z^{\prime}$ boson, additionally we will derive new ones related to the scalars $S_2$ and $I_3$.

\section{Neutral currents via scalar bosons exchange}

In this section we will derive the new sources of FCNC coming from the CP-even ($S_2$) and CP-odd ($I^0_3$) neutral scalars by analyzing the Yukawa Lagrangian Eq.\eqref{yukawafermions}. Despite having five neutral scalars fields only three of them develop a nonzero vacuum expectation value ($vev$) to generate mass for all particles. Expanding these fields around their $vev$s we find
\begin{equation}
\label{vacuos1}
\chi^{\prime 0}, \rho^{0}, \eta^{0} \rightarrow \dfrac{1}{\sqrt{2}}(v_{\chi^{\prime}, \rho, \eta}+R_{\chi^{\prime}, \rho, \eta}+ iI_{\chi', \rho, \eta}).
\end{equation}

Substituting Eq.\eqref{vacuos1} into Eq.\eqref{yukawafermions},  we obtain the mass matrix for the standard down-quarks in the flavor basis ($d_{1}$, $d_{2}$, $d_{3}$),
\begin{equation}
\label{massV}
M^{D}=
\dfrac{1}{\sqrt{2}}
\begin{pmatrix}
\lambda_{411}v & \lambda_{412}v & \lambda_{413}v \\
\lambda_{421}v & \lambda_{422}v & \lambda_{423}v \\
\lambda_{11}v & \lambda_{12}v & \lambda_{13}v
\end{pmatrix},
\end{equation}

as well as the standard up-quarks and exotic ones in the flavor basis ($u_{1}$, $u_{2}$, $u_{3}$) and ($\textsc{u}^{\prime}_{3}$, $\textsc{d}^{\prime}_{1}$, $\textsc{d}^{\prime}_{2}$) respectively,

\begin{equation}
\label{massV2}
M^{U}=\dfrac{1}{\sqrt{2}}
\begin{pmatrix}
-\lambda_{211}v & -\lambda_{212}v & -\lambda_{213}v \\
-\lambda_{221}v & -\lambda_{222}v & -\lambda_{223}v \\
\lambda_{31}v & \lambda_{32}v & \lambda_{33}v
\end{pmatrix},
\end{equation}

\begin{equation}
\label{massUD}
M^{\textsc{d}^{\prime}}_{\textsc{u}^{\prime}}=
\dfrac{1}{\sqrt{2}}
\begin{pmatrix}
\lambda_{1}v_{\chi^{\prime}} & 0 & 0 \\
0 & \lambda_{211}v_{\chi^{\prime}} & \lambda_{212}v_{\chi^{\prime}} \\
0 & \lambda_{221}v_{\chi^{\prime}} & \lambda_{222}v_{\chi^{\prime}}
\end{pmatrix},
\end{equation}
where $\lambda^{\prime}$s refers to the Yukawa coupling constants defined in Eq.\eqref{yukawafermions}.

We can clearly check from Eqs.\eqref{massV}-\eqref{massUD} that the standard quarks do not mix with the exotic ones, justifying our transformations given in Eq.\eqref{transformationL}. This conclusion could be different if we had allowed the neutral scalar $\eta_{0}^{\prime}$ develop a non zero vev. This scenario would lead to mixing among the $W^{\pm}$ and the $V^{\pm}$ and consequently to changes in the W bosons couplings with standard model particles which are largely disfavored by the precise measurements regarding the $W^{\pm}$ properties and couplings\cite{PDGW}.

With these matrices we can find the Yukawa Lagrangian Eq.\eqref{yukawafermions} in terms of the physical scalar bases given determined in Eq.(\ref{Hdfisicos1})-(\ref{Hufisicos3}) in the appendix(\ref{appendix}). Through these, we notice that after substituting the transformations given in Eq.(\ref{misturaquarksMP}) the standard Higgs boson does not mediate FCNC processes, while the physical scalars $S_{2}$ and $I^{0}_{3}$ might mediate, because their interactions with the physical Standard quarks are not flavor diagonal.

In order to estimate which terms in Eq.\eqref{yukawafermions} induce the meson oscillations $K^{0}-\bar{K^{0}}$, $D^{0}-\bar{D^{0}}$ and $B^{0}_d-\bar{B^{0}_d}$  we have used a parametrization Fritzsch type ~\cite{fritzsch} with 4 texture zeros described in \cite{PRDColombia}.  Hereupon we present all terms which contribute to FCNC in the scalar sector,

\begin{eqnarray}
 & &\mathcal{L}^{K^{0}-\bar{K^{0}}}_{S_2,I_3} =  \{ \frac{\lambda_{413}}{2}(V^d_L)^{\ast}_{11}(V^d_R)_{32}+ \frac{\lambda_{423}}{2} (V^d_L)^{\ast}_{21}(V^d_R)_{32}  \nonumber \\
 & & -\frac{\lambda_{13}}{2}(V^d_L)^{\ast}_{31}(V^d_R)_{32}+ \frac{\lambda_{422}}{2} (V^d_L)^{\ast}_{21}(V^d_R)_{22}  \nonumber \\
 & &  -\frac{\lambda_{12}}{2}(V^d_L)^{\ast}_{31}(V^d_R)_{22} - \frac{\lambda_{11}}{2} (V^d_L)^{\ast}_{31}(V^d_R)_{12} \} [\bar{d^{\prime}_{1L}} d^{\prime}_{2R}]   \ \nonumber \\
 & &  \left(  S_2,I_3 \right) .
 \label{FCNC4}
\end{eqnarray}

\begin{eqnarray}
 & &\mathcal{L}^{D^{0}-\bar{D^{0}}}_{S_2,I_3} = \{ \frac{\lambda_{31}}{2}(V^u_L)^{\ast}_{31}(V^u_R)_{12}+\frac{\lambda_{222}}{2}(V^u_L)^{\ast}_{21}(V^u_R)_{22}  \nonumber \\
 & & +\frac{\lambda_{32}}{2}(V^u_L)^{\ast}_{31}(V^u_R)_{22}+ \frac{\lambda_{213}}{2} (V^u_L)^{\ast}_{11}(V^u_R)_{32}  \nonumber \\
 & &  +\frac{\lambda_{223}}{2} (V^u_L)^{\ast}_{21}(V^u_R)_{32}\nonumber+ \frac{\lambda_{33}}{2} (V^u_L)^{\ast}_{31}(V^u_R)_{32} \}[\bar{u^{\prime}_{1L} }u^{\prime}_{2R}] \\
 & & (S_2,I_3).
 \label{FCNC5}
\end{eqnarray}

\begin{eqnarray}
 & &\mathcal{L}^{B^{0}_d-\bar{B^{0}_d}}_{S_2,I_3} =  \{ \frac{\lambda_{413}}{2}(V^d_L)^{\ast}_{11}(V^d_R)_{33}+ \frac{\lambda_{423}}{2} (V^d_L)^{\ast}_{21}(V^d_R)_{33}  \nonumber \\
 & & -\frac{\lambda_{13}}{2}(V^d_L)^{\ast}_{31}(V^d_R)_{33}+ \frac{\lambda_{422}}{2} (V^d_L)^{\ast}_{21}(V^d_R)_{23}  \nonumber \\
 & &  -\frac{\lambda_{12}}{2}(V^d_L)^{\ast}_{31}(V^d_R)_{23} - \frac{\lambda_{11}}{2} (V^d_L)^{\ast}_{31}(V^d_R)_{13} \} [\bar{d^{\prime}_{1L}} d^{\prime}_{3R}]   \ \nonumber \\
 & &  \left(  S_2,I_3 \right) .
 \label{FCNC6}
\end{eqnarray}

All parameters which enter in the Eqs.(\ref{FCNC1})-(\ref{FCNC6}) are known and given in Appendix \ref{appendix3}. In particular the Yukawa parameters which appear in the Eqs.(\ref{FCNC4})-(\ref{FCNC6}) are determined by the quarks' masses, and all matrix elements are constrained by the CKM matrix. We obtained all of them by comparing our mass matrices in the Eqs.(\ref{massV})-(\ref{massV2}) with the ones found in \cite{PRDColombia}. We may notice that the above expressions give us the relations among the scalars and the mixing matrices elements. In other words, they provide the new sources of FCNC in the $331_{RHN}$, that we will explore further.

% for when $M_{S2}=M_{I_{0}^{3}}$, which is plausible because its masses depend upon the same vev, $v_{\chi^{\prime}}$

\section{Meson mixing at tree level}

In this section we are going to find the scalars $S_2$, $I_3$ and the $Z^{\prime}$ boson contributions at tree level to the mass difference system of the mesons systems $K_{0}-\bar{K_{0}}$ and $D^{0}-\bar{D^{0}}$ and $B^{0}_d-\bar{B^{0}_d}$. It is straightforward from Eqs.\eqref{FCNC1}-\eqref{FCNC3} to get the respective effective Lagrangians,

\begin{eqnarray}
\ensuremath{\mathcal{L}}^{K_{0}-\bar{K_{0}}}_{Z'\ eff} & = & \frac{4 \sqrt{2} G_F C^4_W}{(3-4S^2_W)}\frac{M_{Z}^{2}}{M_{Z^{\prime}}^{2}}|(V_{L}^{d})_{31}^{\ast}(V_{L}^{d})_{32}|^{2}|\bar{d}_{1L}^{\prime}\gamma_{\mu}d_{2L}^{\prime}|^{2}\nonumber \\,
\label{effectiveLK}
\end{eqnarray}

\begin{eqnarray}
\ensuremath{\mathcal{L}}^{D_{0}-\bar{D_{0}}}_{Z'\ eff} & = & \frac{4 \sqrt{2} G_F C^4_W}{(3-4S^2_W)}\frac{M_{Z}^{2}}{M_{Z^{\prime}}^{2}}|(V_{L}^{u})_{31}^{\ast}(V_{L}^{u})_{32}|^{2}|\bar{u}_{1L}^{\prime}\gamma_{\mu}u_{2L}^{\prime}|^{2}\nonumber \\,
\label{effectiveLD}
\end{eqnarray}

\begin{eqnarray}
\ensuremath{\mathcal{L}}^{B^{0}_d-\bar{B^{0}_d}}_{Z'\ eff} & = & \frac{4 \sqrt{2} G_F C^4_W}{(3-4S^2_W)}\frac{M_{Z}^{2}}{M_{Z^{\prime}}^{2}}|(V_{L}^{d})_{31}^{\ast}(V_{L}^{d})_{33}|^{2}|\bar{d}_{1L}^{\prime}\gamma_{\mu}d_{3L}^{\prime}|^{2}\nonumber \\,
\label{effectiveLB}
\end{eqnarray}

These effective Lagrangians are in perfect agreement with previous works \cite{PRDColombia} in the limit that $Z_1 \equiv Z$ and $Z_2\equiv Z^{\prime}$ as we are assuming here, and from them, we inherit the subsequently mass difference terms,

\begin{eqnarray}
(\Delta m_{K})_{Z^{\prime}} & = &\frac{4 \sqrt{2} G_F C^4_W}{(3-4S^2_W)}\frac{M_{Z}^{2}}{M_{Z^{\prime}}^{2}}|(V_{L}^{d})_{31}^{\ast}(V_{L}^{d})_{32}|^{2}f_{K}^{2}B_{K}\eta_{K}m_{k},\nonumber \\
\label{massdifKZlinha}
\end{eqnarray}

\begin{eqnarray}
(\Delta m_{D})_{Z^{\prime}} & = &\frac{4 \sqrt{2} G_F C^4_W}{(3-4S^2_W)}\frac{M_{Z}^{2}}{M_{Z^{\prime}}^{2}}|(V_{L}^{u})_{31}^{\ast}(V_{L}^{u})_{32}|^{2}f_{D}^{2}B_{D}\eta_{D}m_{D},\nonumber \\
\label{massdifDZlinha}
\end{eqnarray}

\begin{eqnarray}
(\Delta m_{B_d})_{Z^{\prime}} & = &\frac{4 \sqrt{2} G_F C^4_W}{(3-4S^2_W)}\frac{M_{Z}^{2}}{M_{Z^{\prime}}^{2}}|(V_{L}^{d})_{31}^{\ast}(V_{L}^{d})_{33}|^{2}f_{B}^{2}B_{B}\eta_{B}m_{B},\nonumber \\
\label{massdifBZlinha}
\end{eqnarray}

Here $B$ and $f$ are the bag parameters and the decay constant of the mesons respectively, and $\eta$ the leading order QCD corrections~\cite{etaparameter}. 
We will be using the numerical values $G_F= 1.166 \times 10^{-5}\ \mbox{GeV}^{-2}, (\Delta m_{K}) = 3.483 \times 10^{-12}\ \mbox{MeV},m_{K} = 497.614\ \mbox{MeV},\sqrt{B_{K}}f_{K} = 135\ \mbox{MeV},\eta_K  = 0.57; (\Delta m_{D}) = 4.607\times 10^{-11}\ \mbox{MeV},m_{D}=1865\ \mbox{MeV}, \sqrt{B_{D}}f_{D} =187\ \mbox{MeV}, \eta_D=0.57;(\Delta m_{B_d}) = 3.33\times 10^{-10}\ \mbox{MeV},m_{B_d} = 5279.5 \ \mbox{MeV},
\sqrt{B_{B_d}}f_{B_d} = 208\ \mbox{MeV}, \eta_{B_d} = 0.55$, according to \cite{PDG}.

The mass difference terms associated with the scalar Lagrangians Eq.\eqref{FCNC4}-\eqref{FCNC6} are estimated using the approach described in ~\cite{moha} which is similar to what we have done in the case of the $Z^{\prime}$ boson above. Using this procedure, from Eqs.(\ref{FCNC4})-(\ref{FCNC6}) we find the new terms which contribute to the mass difference terms of the mesons system,

\begin{eqnarray}
(\Delta m_{K})_{S2,I_{0}^{3}} & = & \frac{A_1}{4M_{S_{2},I_{3}^{0}}^{2}}\frac{m_{K}^{3}f_{k}^{2}}{(m_{d}+m_{s})^{2}},\nonumber \\
\label{massdifKscalars}
\end{eqnarray}

\begin{eqnarray}
(\Delta m_{D})_{S2,I_{0}^{3}} & = & \frac{A_2}{4M_{S_{2},I_{3}^{0}}^{2}}\frac{m_{D}^{3}f_{D}^{2}}{(m_{u}+m_{c})^{2}},\nonumber \\
\label{massdifDscalars}
\end{eqnarray}

\begin{eqnarray}
(\Delta m_{B_d})_{S2,I_{0}^{3}} & = & \frac{A_3}{4M_{S_{2},I_{3}^{0}}^{2}}\frac{m_{B}^{3}f_{B}^{2}}{(m_{d}+m_{b})^{2}},\nonumber \\
\label{massdifBscalars}
\end{eqnarray}

where $A_1$, $A_2$ and $A_3$ are the numeric value that we find after summing up all coefficients between curly brackets in the Eq.(\ref{FCNC4})-(\ref{FCNC6}) and squaring respectively, and $m_u, m_d, m_c, m_s$ and $m_b$ are the standard quarks' masses.

Now we are going to show our results using the Eqs.(\ref{massdifKZlinha})-(\ref{massdifBscalars}) which represent all contributions coming from the model to the mesons oscillations systems in the study. These can be rewritten in terms of the mass of the mediators only, after plugging in all parameters and constants. The final equations are presented in the Appendix \ref{appendix3} through the Eqs.(\ref{finalEq1}-\ref{finalEq6}). In our analyses we implemented the current constraints on the Higgs mass ($M_H\simeq 120$ GeV) according to recent LHC and TEVATRON observations \cite{LHChiggs}, as well as one related to the $Z^{\prime}$ search \cite{limitZlinhaLHC} ($M_{Z^{\prime}} \gtrsim 1.6$TeV) and the measurements on the mass difference of the mesons \cite{PDG}.

First of all, we would like to point out that through the Eqs.\eqref{psescalaresfisicos}-\eqref{massvec} we recognize that $I_3$ and $Z^{\prime}$ masses are determined just by $v_{\chi^{\prime}}$ and  when we fix the Higgs mass, automatically the mass of $S_2$ also depends on $v_{\chi^{\prime}}$ only. Thereupon our results rest on one free parameter only and as a result we can set strong constraints on the parameter space. Although one can discern that the $Z^{\prime}$ contribution is the most relevant one for mesons oscillations purposes, scalars contributions play a role also in this processes and therefore it is worthwhile to investigate them and derive new bounds on space of parameters.

\begin{figure}[!htb]
\includegraphics[scale=0.6]{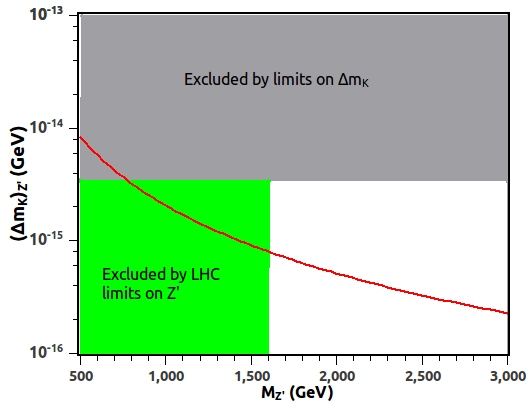}
\caption{The gray region is the excluded region defined by ($\Delta m_K$). The green region reflects the constraints on $M_Z^{\prime} \geq 1.6$ TeV reported by CMS and ATLAS \cite{limitZlinhaLHC}. This CMS and ATLAS limit may not apply for the case of this 331 model though. We are being conservative and plotting this constraint anyway.The red curve is the $Z^{\prime}$ contribution to ($\Delta m_K$) respectively. So $K^0-\bar{K^0}$ bound implies that $M_{Z^{\prime}} \gtrsim 770$~GeV.See text for more details}
\label{fig1}
\end{figure}

\begin{figure}[!htb]
\includegraphics[scale=0.6]{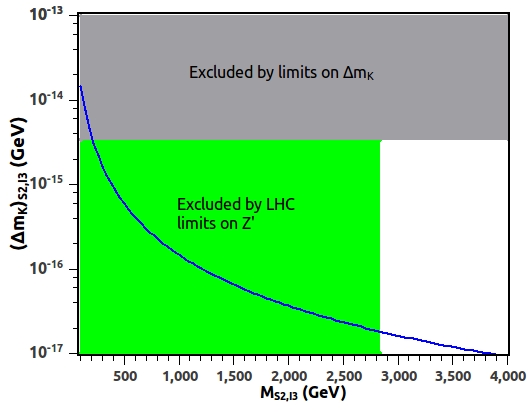}
\caption{The gray region is the excluded region defined by ($\Delta m_K$). The green region reflects the constraints on $M_Z^{\prime} \geq 1.6$ TeV reported by CMS and ATLAS \cite{limitZlinhaLHC}. This CMS and ATLAS limit may not apply for the case of this 331 model though.  We are being conservative and plotting this constraint anyway. The blue curve is the $S_2$ and $I_3$ contribution to ($\Delta m_K$) in the model.  $S_2$ and $I_3$ have equal contributions to ($\Delta m_K$) for this reason we showed only one curve for both. Considering only this contribution we set the limit $M_{S_2,I_3} \gtrsim 200$~GeV. Taking into account all contributions we find $M_{Z^{\prime}} \gtrsim 770$~GeV and  $M_{S_2},M_{I_3} \gtrsim 1376$~GeV. See text for more details}
\label{fig2}
\end{figure}

In FIG.\ref{fig1} we exhibit $\Delta m_{K}$ in terms of the $Z^{\prime}$ mass and in FIG.\ref{fig2} as function of the masses of $S_2$ and $I_3$. We distinctly observe that the $Z^{\prime}$ contribution is the most important one while the scalars ones are suppressed. For this reason we can set stronger constraints on the mass of $Z^{\prime}$. Indeed, in order to $(\Delta m_{K}) \leqslant 3.483 \times 10^{-15}$~GeV from FIG.\ref{fig1} we find that $M_{Z^{\prime}} \gtrsim 770$ GeV and from FIG.\ref{fig2} $M_{S_2,I_3} \gtrsim 200$~GeV. Nevertheless as aforementioned, the masses of $Z^{\prime}$ and the scalars depend only on $v_{\chi^{\prime}}$ thus the requirement $M_{Z^{\prime}} \gtrsim 770$ implies that $v_{\chi^{\prime}} \gtrsim 1945$~GeV which imposes that $M_{S_2},M_{I_3} \gtrsim 1376$~GeV. Therefore the latter is the bound that we get from $K^0-\bar{K^0}$ system.

\begin{figure}[!htb]
\includegraphics[scale=0.6]{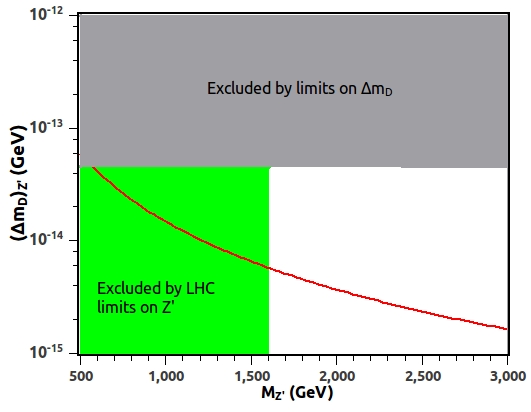}
\caption{The gray region is excluded region defined by ($\Delta m_D$). The green region reflects the constraints on $M_Z^{\prime} \geq 1.6$ TeV reported by CMS and ATLAS \cite{limitZlinhaLHC}. This CMS and ATLAS limit may not apply for the case of this 331 model though. We are being conservative and plotting this constraint anyway.The red curve is the $Z^{\prime}$ contribution to ($\Delta m_D$) respectively. So $D^0-\bar{D^0}$ bound implies that $M_{Z^{\prime}} \gtrsim 550$~GeV.See text for more details.}
\label{fig3}
\end{figure}

\begin{figure}[!htb]
\includegraphics[scale=0.6]{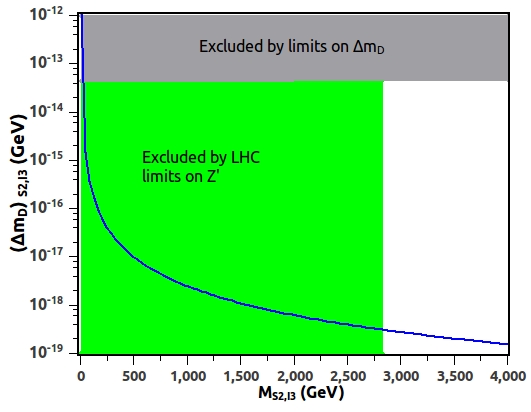}
\caption{The gray region is the excluded region defined by ($\Delta m_D$).The green region reflects the constraints on $M_Z^{\prime} \geq 1.6$ TeV reported by CMS and ATLAS \cite{limitZlinhaLHC}. This CMS and ATLAS limit may not apply for the case of this 331 model though.  We are being conservative and plotting this constraint anyway. The blue curve is the $S_2$ and $I_3$ contribution to ($\Delta m_K$) in the model.  $S_2$ and $I_3$ have equal contributions to ($\Delta m_D$) for this reason we showed only one curve for both. Considering only this contribution we set the limit $M_{S_2,I_3} \gtrsim 1$~GeV. Taking into account all contributions we find $M_{Z^{\prime}} \gtrsim 550$~GeV and  $M_{S_2},M_{I_3} \gtrsim 983$~GeV. See text for more details.}
\label{fig4}
\end{figure}

In FIG.\ref{fig3} the precise measurements on $D^0-\bar{D^0}$, that is the limit $(\Delta m_{D})\leqslant 4.607 \times 10^{-14}$~GeV requires that $M_{Z^{\prime}} \gtrsim 550$~GeV and from FIG.\ref{fig4} we set $M_{S_2},M_{I_3} \gtrsim 1$~GeV. Again we use the fact that these results are correlated and hence $M_{Z^{\prime}} \gtrsim 550$~GeV infer $M_{S_2},M_{I_3} \gtrsim 980$~GeV.

\begin{figure}[!htb]
\includegraphics[scale=0.6]{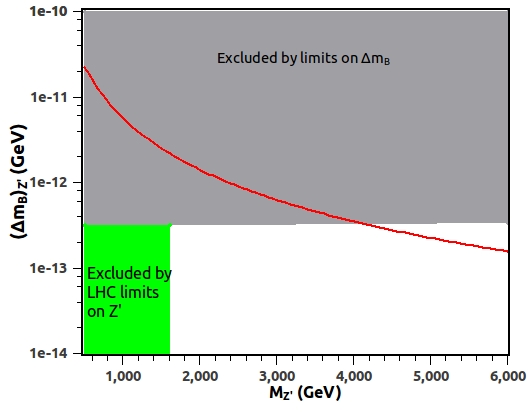}
\caption{The gray region is the excluded region defined by ($\Delta m_{B_d}$). The green region reflects the constraints on $M_Z^{\prime} \geq 1.6$ TeV reported by CMS and ATLAS \cite{limitZlinhaLHC}. This CMS and ATLAS limit may not apply for the case of this 331 model though. We are being conservative and plotting this constraint anyway.The red curve is the $Z^{\prime}$ contribution to ($\Delta m_D$) respectively. So $B^0_d-\bar{B^0_d}$ bound implies that $M_{Z^{\prime}} \gtrsim 4.2$~TeV.See text for more details.}
\label{fig6}
\end{figure}

\begin{figure}[!htb]
\includegraphics[scale=0.6]{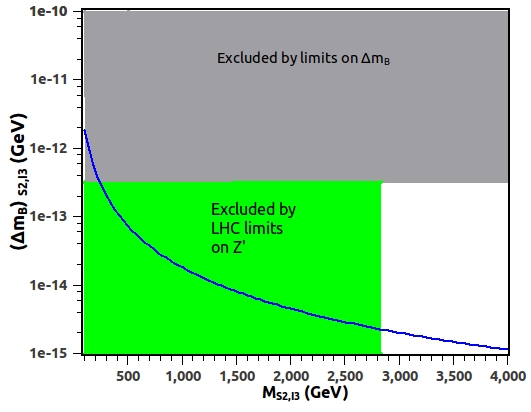}
\caption{The gray region is the excluded region defined by ($\Delta m_{B_d}$). The green region reflects the constraints on $M_Z^{\prime} \geq 1.6$ TeV reported by CMS and ATLAS \cite{limitZlinhaLHC}. This CMS and ATLAS limit may not apply for the case of this 331 model though.  We are being conservative and plotting this constraint anyway. The blue curve is the $S_2$ and $I_3$ contribution to ($\Delta m_{B_d}$) in the model.  $S_2$ and $I_3$ have equal contributions to ($\Delta m_{B_d}$) for this reason we showed only one curve for both. Considering only this contribution we set the limit $M_{S_2,I_3} \gtrsim 230$~GeV. Taking into account all contributions we find $M_{Z^{\prime}} \gtrsim 4.2$~TeV and  $M_{S_2},M_{I_3} \gtrsim 7.5$~TeV. See text for more details.}
\label{fig5}
\end{figure}

In FIG.\ref{fig6}-\ref{fig5} the demand $(\Delta m_{B_d}) \leqslant 3.33 \times 10^{-13}$~GeV entreats that $M_Z^{\prime} \gtrsim 4.2$ TeV which implies in $v_\chi^{\prime} \gtrsim 10.6$ TeV and consequently $M_{S_2},M_{I_3} \gtrsim 7.5$ TeV. We can notice in FIG.\ref{fig2} that this limit is even stronger than the LHC one in the mass of $Z^{\prime}$. Hence the precise measurement on $B^0_d-\bar{B^0_d}$ oscillations rule out a large region of the parameter space of the model, and it truly makes the detection of the $Z^{\prime}$ of the $331_{RHN}$ very unlikely in the current LHC energy range.

In summary the strongest constraint on the model comes from the precise measurements on the $B^0_d-\bar{B^0_d}$ system which demands that  $M_Z^{\prime} \gtrsim 4.2$ TeV and $M_{S_2},M_{I_3} \gtrsim 7.5$ TeV.

\vspace{20cm}

\section{Conclusions}

In this work we have shown that the $Z^{\prime}$ gauge boson is not the unique source of FCNC in the $331_{RHN}$. Instead we have two new contributions coming from the CP-even and -odd scalar $S_2$ and $I_3$. These account for all possible FCNC processes in the model. We have also built analytical expressions for the mass difference of the meson systems $K_{0}-\bar{K_{0}}$, $D^{0}-\bar{D^0}$, $B^0_d-\bar{B^0_d}$ taking into account all terms and assuming a 4 texture zeros approach. It is important to emphasize that in more general setups regarding the quarks mass matrices others contributions may arise affecting our conclusions.

First, we have shown explicitly that, in the limit case $\Phi=0$, we recover the flavor diagonal interactions of the $Z$ boson with standard quarks and obtain a flavor non-diagonal interaction of them with the new neutral $Z^{\prime}$ boson, which contributes at tree level to FCNC processes and consequently to mass difference terms in perfect agreement with previous works.

In addition we have derived the new ones coming from the scalar namely $S_2$ and $I_3$, and included the current constraints on the Higgs and $Z^{\prime}$ masses and the bounds on the mass differences of these mesons.

Our results are summarized in FIG.\ref{fig1}-\ref{fig5} and they rest on one parameter only. Through them we discerned that the $Z^{\prime}$ contributions are the most relevant one for mesons oscillations purposes, we also could be able to strengthen constraints on the masses of the mediators, and in particular, the limits on $B^0_d-\bar{B^0_d}$ system demand that $M_Z^{\prime} \gtrsim 4.2$ TeV and $M_{S_2},M_{I_3} \gtrsim 7.5$ TeV which renders the detection of the $Z^{\prime}$ of the $331_{RHN}$ very unlikely in the current LHC energy range.

\appendix
\section{}
\label{appendix1}

The lagrangian that leads us to the FCNC phenomena mediated by the $Z^2$ boson, which contributes at tree level to the mass difference of the neutral meson systems, is:
\begin{equation}
\label{lagrangianaquarks331}
\ensuremath{\mathcal{L}}^{331_{RH_{\nu}}}_{FCNC}= [\bar{Q}_{L}^{3}i\gamma^{\mu}D^{L}_{\mu}Q_{L}^{3}+\sum_{i=1}^{2}\bar{Q}_{L}^{i}i\gamma^{\mu}D^{L\star}_{\mu}Q_{L}^{i}],
\end{equation}
being $D_{\mu}^{L} = \partial_{\mu}+\frac{1}{2} igW_{\mu}^{a} \lambda^{a}+i\ {\bf g_{N}} N \omega_{\mu}^{N}$, the covariant derivate for triplets, $W_{\mu}^{a}$ the symmetrical gauge bosons of the $SU(3)_{L}$ group, $B_{\mu}$ the symmetrical gauge boson of the $U(1)_{N}$ group, and $\lambda^{a}$ the Gell-Mann matrices. Since symmetrical gauge bosons are different from the physical ones we must diagonalize their mass matrices.

In terms of the physical neutral gauge bosons we rewrite \eqref{lagrangianaquarks331} as
\begin{eqnarray}
\bar{Q}_{iL}i\gamma^{\mu}D_{\mu}^{L}Q_{iL} & = & g\{J_{z^1}^1\bar{d}_{iL}\gamma^{\mu}d_{iL}Z_{\mu}^{1}+J_{z^2}^1\bar{d}_{iL}\gamma^{\mu}d_{iL}Z_{\mu}^{2}\nonumber\\
                                           &   &
+J_{z^1}^2\bar{u}_{iL}\gamma^{\mu}u_{iL}Z_{\mu}^{1}+J_{z^2}^2\bar{u}_{iL}\gamma^{\mu}u_{iL}Z_{\mu}^{2}\nonumber\\
                                           &   &
+\dfrac{S_{W}}{3}\bar{d}_{iL}\gamma^{\mu}d_{iL} A_{\mu}-\dfrac{2S_{W}}{3}\bar{u}_{iL}\gamma^{\mu}u_{iL}A_{\mu}\nonumber\\
                                           &   &
+J_{z^1}^3\bar{\textsc{d}}^{\prime}_{iL}\gamma^{\mu}\textsc{d}^{\prime}_{iL}Z_{\mu}^{1}+J_{z^2}^3\bar{\textsc{d}}^{\prime}_{iL}\gamma^{\mu}\textsc{d}^{\prime}_{iL}Z_{\mu}^{2}\nonumber\\
                                           &   &
+\dfrac{S_{W}}{3}\bar{\textsc{d}}^{\prime}_{iL}\gamma^{\mu}\textsc{d}^{\prime}_{iL}A_{\mu}\},\nonumber\\
\label{neutralquark1}
\end{eqnarray}

with

\begin{eqnarray}
J_{z^1}^1 & = & \dfrac{C_{W}C_{\Phi}}{2}+\dfrac{t_{W}S_{W}C_{\Phi}}{6}+\dfrac{\sqrt{h_{W}}S_{\Phi}}{6C_{W}},\nonumber\\
J_{z^2}^1 & = & -\dfrac{C_{W}S_{\Phi}}{2}+\dfrac{\sqrt{h_{W}}C_{\Phi}}{6C_{W}}-\dfrac{t_{W}S_{W}S_{\Phi}}{6},\nonumber\\
J_{z^1}^2 & = & -\dfrac{C_{W}C_{\Phi}}{2}+\dfrac{t_{W}S_{W}C_{\Phi}}{6}+\dfrac{\sqrt{h_{W}}S_{\Phi}}{6C_{W}},\nonumber\\
J_{z^2}^2 & = & \dfrac{C_{W}S_{\Phi}}{2}+\dfrac{\sqrt{h_{W}}C_{\Phi}}{6C_{W}}-\dfrac{t_{W}S_{W}S_{\Phi}}{6},\nonumber\\
J_{z^1}^3 & = & -\dfrac{\sqrt{h_{W}}S_{\Phi}}{3C_{W}}-\dfrac{t_{W}S_{W}C_{\Phi}}{3},\nonumber\\
J_{z^2}^3 & = & -\dfrac{\sqrt{h_{W}}C_{\Phi}}{3C_{W}}+\dfrac{t_{W}S_{W}S_{\Phi}}{3}.
\end{eqnarray}

and

\begin{eqnarray}
\bar{Q}_{3L}i\gamma^{\mu}D_{\mu}^{L}Q_{3L} & = & -g\{K_{z^1}^1\bar{u}_{3L}\gamma^{\mu}u_{3L}Z_{\mu}^{1}+K_{z^2}^1\bar{u}_{3L}\gamma^{\mu}u_{3L}Z_{\mu}^{2}\nonumber\\
                                           &   &
+K_{z^1}^2\bar{d}_{3L}\gamma^{\mu}d_{3L}Z_{\mu}^{1}+K_{z^2}^2\bar{d}_{3L}\gamma^{\mu}d_{3L}Z_{\mu}^{2}\nonumber\\
                                           &   &
+\dfrac{2}{3}S_{W}\bar{u}_{3L}\gamma^{\mu}u_{3L} A_{\mu}-\dfrac{S_{W}}{3}\bar{d}_{3L}\gamma^{\mu}d_{3L}A_{\mu}\nonumber\\
                                           &   &
+K_{z^1}^3\bar{\textsc{u}}^{\prime}_{3L}\gamma^{\mu}\textsc{u}^{\prime}_{3L}Z_{\mu}^{1}+K_{z^2}^3\bar{\textsc{u}}^{\prime}_{3L}\gamma^{\mu}\textsc{u}^{\prime}_{3L}Z_{\mu}^{2}\nonumber\\
                                           &   &
+\dfrac{2}{3}S_{W}\bar{\textsc{u}}^{\prime}_{3L}\gamma^{\mu}\textsc{u}^{\prime}_{3L}A_{\mu}\},\nonumber\\
\label{neutralquark3}
\end{eqnarray}

with
\begin{eqnarray}
K_{z^1}^1 & = & \dfrac{C_{W}C_{\Phi}}{2}-\dfrac{t_{W}S_{W}C_{\Phi}}{2}+\dfrac{\sqrt{h_{W}}S_{\Phi}}{6C_{W}}+\dfrac{t_{W}S_{W}S_{\Phi}}{3\sqrt{h_{W}}},\nonumber\\
K_{z^2}^1 & = & -\dfrac{C_{W}S_{\Phi}}{2}+\dfrac{\sqrt{h_{W}}C_{\Phi}}{6C_{W}}+\dfrac{t_{W}S_{W}S_{\Phi}}{2}+\dfrac{t_{W}S_{W}C_{\Phi}}{3\sqrt{h_{W}}},\nonumber\\
K_{z^1}^2 & = & -\dfrac{C_{W}C_{\Phi}}{2}-\dfrac{t_{W}S_{W}C_{\Phi}}{6}+\dfrac{\sqrt{h_{W}}S_{\Phi}}{6C_{W}}+\dfrac{t_{W}S_{W}S_{\Phi}}{3\sqrt{h_{W}}},\nonumber\\
K_{z^2}^2 & = & \dfrac{C_{W}S_{\Phi}}{2}+\dfrac{\sqrt{h_{W}}C_{\Phi}}{6C_{W}}+\dfrac{t_{W}S_{W}S_{\Phi}}{6}+\dfrac{t_{W}S_{W}C_{\Phi}}{3\sqrt{h_{W}}},\nonumber\\
K_{z^1}^3 & = & -\dfrac{\sqrt{h_{W}}S_{\Phi}}{3C_{W}}+\dfrac{t_{W}S_{W}S_{\Phi}}{3\sqrt{h_{W}}}-\dfrac{2}{3}t_{W}S_{W}C_{\Phi},\nonumber\\
K_{z^2}^3 & = & -\dfrac{\sqrt{h_{W}}C_{\Phi}}{3C_{W}}+\dfrac{t_{W}S_{W}C_{\Phi}}{3\sqrt{h_{W}}}+\dfrac{2}{3}t_{W}S_{W}S_{\Phi},
\end{eqnarray}
\vspace{0.5cm}

where $S_{W}$ is the sine of the Weinberg angle, $S_{\Phi}$ is the sine of the mixing angle between the physical gauge bosons $Z^{1}$ and $Z^{2}$, $g$ is the $SU(2)_{L}$ coupling constant, $h_{W}=3-4S_{W}^2$, and $A_{\mu}$ the massless physical gauge boson of the theory identified as the photon. Finally taking $\Phi=0$ in Eqs.\eqref{neutralquark1} and \eqref{neutralquark3} we obtain Eqs.\eqref{cnLRquarks}-\eqref{zprimad}.

\vspace{0.5cm}

\section{}
\label{appendix}

\begin{equation}
\label{Hdfisicos1}
\ensuremath{\mathcal{L}}_{d}=
\frac{1}{2}
\begin{pmatrix}
\bar{d}^{\prime}_{1L} & \bar{d}^{\prime}_{2L} & \bar{d}^{\prime}_{3L}
\end{pmatrix}
\begin{pmatrix}
\lambda_{411} & \lambda_{412} & \lambda_{413} \\
\lambda_{421} & \lambda_{422} & \lambda_{423} \\
\lambda_{11}  & \lambda_{12}  & \lambda_{13}
\end{pmatrix}
\begin{pmatrix}
d^{\prime}_{1R} \\
d^{\prime}_{2R} \\
d^{\prime}_{3R}
\end{pmatrix}H,
\end{equation}

\begin{equation}
\label{Hdfisicos2}
+\frac{1}{2}
\begin{pmatrix}
\bar{d}^{\prime}_{1L} & \bar{d}^{\prime}_{2L} & \bar{d}^{\prime}_{3L}
\end{pmatrix}
\begin{pmatrix}
\lambda_{411} & \lambda_{412} & \lambda_{413} \\
\lambda_{421} & \lambda_{422} & \lambda_{423} \\
-\lambda_{11} & -\lambda_{12} & -\lambda_{13}
\end{pmatrix}
\begin{pmatrix}
d^{\prime}_{1R} \\
d^{\prime}_{2R} \\
d^{\prime}_{3R}
\end{pmatrix}S_{2},
\end{equation}

\begin{equation}
\label{Hdfisicos2}
+\frac{1}{2}
\begin{pmatrix}
\bar{d}^{\prime}_{1L} & \bar{d}^{\prime}_{2L} & \bar{d}^{\prime}_{3L}
\end{pmatrix}i
\begin{pmatrix}
-\lambda_{411} & -\lambda_{412} & -\lambda_{413} \\
-\lambda_{421} & -\lambda_{422} & -\lambda_{423} \\
\lambda_{11}  & \lambda_{12}  & \lambda_{13}
\end{pmatrix}
\begin{pmatrix}
d^{\prime}_{1R} \\
d^{\prime}_{2R} \\
d^{\prime}_{3R}
\end{pmatrix}I_{3}^{0}.
\end{equation}

\begin{equation}
\label{Hufisicos1}
\ensuremath{\mathcal{L}}_{u}=
\dfrac{1}{2}
\begin{pmatrix}
\bar{u}^{\prime}_{1L} & \bar{u}^{\prime}_{2L} & \bar{u}^{\prime}_{3L}
\end{pmatrix}
\begin{pmatrix}
\lambda_{211} & \lambda_{212} & \lambda_{213} \\
\lambda_{221} & \lambda_{222} & \lambda_{223} \\
\lambda_{31}  & \lambda_{32}  & \lambda_{33}
\end{pmatrix}
\begin{pmatrix}
u^{\prime}_{1R} \\
u^{\prime}_{2R} \\
u^{\prime}_{3R}
\end{pmatrix}S_2,
\end{equation}

\begin{equation}
\label{Hufisicos2}
+\frac{1}{2}
\begin{pmatrix}
\bar{u}^{\prime}_{1L} & \bar{u}^{\prime}_{2L} & \bar{u}^{\prime}_{3L}
\end{pmatrix}
\begin{pmatrix}
-\lambda_{211} & -\lambda_{212} & -\lambda_{213} \\
-\lambda_{221} & -\lambda_{222} & -\lambda_{223} \\
\lambda_{31} & \lambda_{32} & \lambda_{33}
\end{pmatrix}
\begin{pmatrix}
u^{\prime}_{1R} \\
u^{\prime}_{2R} \\
u^{\prime}_{3R}
\end{pmatrix}H,
\end{equation}

\begin{equation}
\label{Hufisicos3}
+\frac{1}{2}
\begin{pmatrix}
\bar{u}^{\prime}_{1L} & \bar{u}^{\prime}_{2L} & \bar{u}^{\prime}_{3L}
\end{pmatrix}i
\begin{pmatrix}
\lambda_{211} & \lambda_{212} & \lambda_{213} \\
\lambda_{221} & \lambda_{222} & \lambda_{223} \\
\lambda_{31} & \lambda_{32} & \lambda_{33}
\end{pmatrix}
\begin{pmatrix}
u^{\prime}_{1R} \\
u^{\prime}_{2R} \\
u^{\prime}_{3R}
\end{pmatrix}I^{0}_{3}.
\end{equation}

\section{}
\label{appendix3}

In this section we exhibit the values of all couplings used to obtain our results. Values for Yukawa parameters used:

\begin{eqnarray}
\lambda_{411} & = & \lambda_{412}=\lambda_{421}=0;\nonumber \\
\lambda_{413} & = & 1.06\times 10^{-3}, \lambda_{422}  = -2.19\times 10^{-4}; \nonumber \\
\lambda_{423} & = & 2.12 \times 10^{-3}, \lambda_{11}  = 1.06 \times 10^{-3};\nonumber \\
\lambda_{12}  & = & 2.19 \times 10^{-4}, \lambda_{13}  = 2.33 \times 10^{-2}.
\end{eqnarray}

Now we show the values of the parameter expressed in the mass difference terms mediated by the scalar,
summarized in the Eqs.(\ref{massdifKscalars}-\ref{massdifBscalars}).

\begin{eqnarray}
A_1 & = & 26.1121 \times 10^{-10};A_2=49\times 10^{-10};\nonumber \\
A_3 & = & 23.1361\times 10^{-8}.
\end{eqnarray}

Hereunder we present the values used in the CKM matrices,

\begin{equation}
V^{u}_{L}=V^{u}_{R}= \left( \begin{array}{ccc}
0.89 & -0.45 & 2.6\times 10^{-2} \\
-0.45 & -0.89 & 5.4\times 10^{-2} \\
4.6\times 10^{-4} & 6\times 10^{-2} & 1 \end{array} \right)
\end{equation}

\begin{equation}
V^{d}_{L}=V^{d}_{R}= \left( \begin{array}{ccc}
0.97 & -0.22 & -0.33\times 10^{-2} \\
-0.22 & -0.97 & 5.4\times 10^{-2} \\
-1.7\times 10^{-2} & 5.8\times 10^{-2} & 1 \end{array} \right)
\end{equation}

In order to make clearer what were the exactly equations we used to obtain Fig.(1-6), hereafter we present the final mass difference mass terms after plugging in all parameters,

\begin{equation}
\label{finalEq1}
(\Delta m_{K})_{Z^{\prime}}= \frac{2.066\times 10^{-9}}{M_{Z^{\prime}}^2}\mbox{(GeV)}
\end{equation}

\begin{equation}
\label{finalEq2}
(\Delta m_{K})_{S2,I_{0}^{3}}= \frac{1.47725\times 10^{-10}}{M_{S_2}^2,M_{I_3}^2}\mbox{(GeV)}
\end{equation}

\begin{equation}
\label{finalEq3}
(\Delta m_{D})_{Z^{\prime}}= \frac{1.48657\times 10^{-8}}{M_{Z^{\prime}}^2}\mbox{(GeV)}
\end{equation}

\begin{equation}
\label{finalEq4}
(\Delta m_{D})_{S2,I_{0}^{3}}= \frac{2.53\times 10^{-12}}{M_{S_2}^2,M_{I_3}^2}\mbox{(GeV)}
\end{equation}

\begin{equation}
\label{finalEq5}
(\Delta m_{B_d})_{Z^{\prime}}= \frac{5.66828\times 10^{-6}}{M_{Z^{\prime}}^2}\mbox{(GeV)}
\end{equation}

\begin{equation}
\label{finalEq6}
(\Delta m_{B_d})_{S2,I_{0}^{3}}= \frac{1.8304\times 10^{-8}}{M_{S_2}^2,M_{I_3}^2}\mbox{(GeV)}
\end{equation}

\begin{acknowledgments}
The authors thank Carlos Pires for valuable discussions and comments. FSQ acknowledges the hospitality of the Universidade Federal de Campina Grande during the early stages of this work. This work is supported by Coordena\c{c}\~ao de Aperfei\c{c}oamento de Pessoal de N\'{\i}vel Superior (CAPES).
\end{acknowledgments}

\end{document}